\documentclass[preprint]{aastex}

\newcommand{\msun}{$M_{\odot}$}

\begin{document}
\title{Near-Infrared Light Curves of the Black Hole Binary A0620--00}
\author{Cynthia S. Froning}
\email{froning@stsci.edu}
\affil{Space  Telescope Science Institute, 3700 San Martin Drive,
Baltimore, MD 21218}
\and
\author{Edward L. Robinson}
\email{elr@astro.as.utexas.edu}
\affil{Department of Astronomy, University of Texas at Austin, 
Austin, TX 78712}

\begin{abstract}

We measured the near-infrared orbital light curve of the black hole
binary A0620--00 in 1995 and 1996.  The light curves show an
asymmetric, double-humped modulation with extra emission in the peak
at orbital phase 0.75.  There were no significant changes in the shape
of the light curve over the one-year observation period.  There were
no sharp dips in the light curves nor reversals of the asymmetry
between the two peaks as seen in earlier observations.  The light
curves are well fit by models incorporating ellipsoidal variations
from the mass-losing K-type star plus a beamed bright spot on the
accretion disk around the compact star.  The long-term stability of
the light curve shape rules out superhumps and star spots as sources
of asymmetry when we observed A0620--00.  The ellipsoidal variations
yield a lower limit $i \geq 38\arcdeg$ on the orbital inclination.
The light curves show no eclipse features, which places an upper limit
$i \leq 75\arcdeg$.  This range of inclinations constrains the mass of
the compact object to $3.3 < M_{1} < 13.6$~\msun.  The light curves do
not further constrain the orbital inclination because the contribution
of the accretion disk to the observed flux is unknown.  We argue that
a previous attempt to measure the near-infrared flux from the
accretion disk using the dilution of the $^{12}$CO(2,0) bandhead in
the spectrum of the K star is not reliable because the band strength
depends strongly on surface gravity.

\end{abstract}


\section{Introduction} \label{sec_intro}

The bright X-ray nova A0620--00 (V616 Mon) was discovered when it
erupted in 1975 \citep{elvis1975}.  Observations obtained after its
return to quiescence revealed that A0620--00 is an interacting binary
system with a K star donating mass to a compact star via an accretion
disk \citep{oke1977,mcclintock1983}.  Early fits to the radial
velocity variations of the K star gave a semi-amplitude of $K_{2} =
457$ km s$^{-1}$, which, when combined with the orbital period
($P_{orb} = 7.75$ hr), yielded a mass function $f(M) = 3.17$~\msun.
The mass function equals the minimum dynamical mass of the compact
star.  Since the mass function for A0620--00 is greater than the
theoretical maximum mass for a neutron star, A0620--00 became a strong
candidate for a black hole binary \citep{mcclintock1986}.  Follow-up
observations have refined the velocity semi-amplitude and determined
the mass ratio of the binary: $K_{2} = 433\pm3$ km s$^{-1}$ and $q =
M_{2} / M_{1} = 0.067\pm0.01$ (Marsh, Robinson \& Wood 1994; Orosz et
al.\ 1994 found similar binary parameters).  With these values, the
masses of the stars are $M_{1} = (3.09\pm0.09)\ \sin^{-3}i$~\msun\ and
$M_{2} = (0.21\pm0.04)\ \sin^{-3}i$~\msun, lacking only a
determination of the orbital inclination to be fully determined.

One method for obtaining the orbital inclination of A0620--00 is to
measure the ellipsoidal variations of the Roche-lobe-filling K star.
Two studies of the ellipsoidal variations have resulted in
non-overlapping estimates of the inclination and, consequently,
non-overlapping estimates of the mass of the compact star.  Haswell et
al.\ (1993) fit models including a Roche-lobe-filling star and an
accretion disk to simultaneous UBVR light curves of A0620--00 and
found an inclination range for $q$ = 0.067 of $65\fdg75 \leq i \leq
73\fdg5$.  This corresponds to a compact star mass of $3.40 \leq M_{1}
\leq 4.20$~\msun\ (the compact star masses given here and elsewhere in
this manuscript were calculated by us using the mass function
determination from Marsh, Robinson \& Wood 1994).  \citet{shahbaz1994}
fit a K-band light curve of A0620--00 with a model including only the
lobe-filling star and derived a best-fit inclination of 37$\arcdeg$,
with a range of inclinations $30\arcdeg \leq i \leq 45\arcdeg$ (90\%
confidence limits for $q$ = 0.067; their Fig.\ 2).  This corresponds
to $M_{1}$ = 14.2~\msun, and a range of $8.49 \leq M_{1} \leq
25.4$~\msun.  All authors agree that the compact star should be a
black hole.

The orbital light curves of A0620--00 have shown clear evidence that
the ellipsoidal variations are distorted: the light curve minima and
maxima have varied in relative height and depth, the light curve
minima have on occasion shown sharp features rather than smooth
troughs, and the asymmetry between the two peaks has reversed (Haswell
1996 and sources therein).  The ellipsoidal variations can be
distorted by star spots on the K star; by a non-axisymmetric
distribution of light across the accretion disk, such as the bright
spot where the mass stream impacts the accretion disk; by superhumps
caused by a precessing, elliptical accretion disk; and by dilution of
the star light by non-variable flux from other components of the
binary system, such as an axisymmetric accretion disk.  The presence
of variable sources of flux in the system can confuse attempts to
isolate the ellipsoidal component of the variations, and if a
non-varying flux is present in the binary system but not in the
models, the models will underestimate the true amplitude of the
ellipsoidal variations and, in turn, underestimate the orbital
inclination.  The difference between the derived inclinations for
A0620--00 stems in part from uncertainty about the magnitude of these
distortions.

The light curve of A0620--00 has been monitored at R and I wavelengths
(e.g., Leibowitz, Hemar \& Orio 1998), but few observations have been
published at longer wavelengths.  Since the contribution of the K star
to the observed flux is maximized in the near-infrared, that
wavelength region provides a good window to target the ellipsoidal
modulation.  In order to investigate the remaining uncertainties
concerning the mass of its compact star and to measure the long-term
behavior of its orbital light curve, we have reobserved the orbital
light curve of A0620--00 in the J, H and K bandpasses.  This paper
reports the results of our observations.

\section{Observations and Data Reduction} \label{sec_observ}

We observed A0620--00 on 1995 December 15 -- 17, 1996 January 27 -- 29
and 1996 December 7 -- 12 on the 2.7-m telescope at McDonald
Observatory using ROKCAM, a near-infrared imaging camera
\citep{colome1993}.  In all three observing runs, we observed
extensively in the H filter (1.45 -- 1.85~$\mu$m), where our S/N was
maximized, and we supplemented these data with observations in the
thermal K (2.05 -- 2.4~$\mu$m) in 1996 January and in the J-band (1.1
-- 1.4~$\mu$m) in 1996 December.  The dates and total exposure times
of the observations are summarized in Table~\ref{tab_logobs}.
Individual exposure times were 20 seconds in J and H and 10 seconds in
K, with telescope nods between integrations to sample the variable sky
background.  After subtracting sky and dark current, we calibrated the
images using dome flats (constructed from images taken with the dome
lights off subtracted from images taken with the lights on).  We
aligned the individual frames and coadded them in groups of four, from
which we extracted the instrumental magnitudes of A0620--00 and three
nearby comparison stars.  The positions of the comparison stars
relative to A0620--00 and their NIR colors are given in
Table~\ref{tab_stars}.

The two brighter comparison stars were averaged and used to correct
the observations for seeing and extinction variations.  The third,
fainter star is comparable in brightness to A0620--00 in the
near-infrared.  We used the scatter in its measurements to assign an
uncertainty to the relative photometry in each filter on each
night.  We flux calibrated the H and K data using observations of four
standard stars from 1996 January 27 \citep{elias1982}.  The J-band
observations were flux calibrated using a single standard star
observed on 1996 December 8; the uncertainty in the J-band flux
calibration was estimated by using the same standard star to calculate
a H-band calibration and comparing that result to the calibration from
1996 January 27.  Because of the paucity of K-band observations of the
A0620--00 field, the transformation equations for each filter were fit
with a fixed extinction coefficient \citep{allen1976} and no color
terms.  The mean colors for each observing run are shown in
Table~\ref{tab_colors}.
%; the error bars given are the uncertainties in
%the absolute flux calibration only.  
The mean H magnitude of the A0620--00 lightcurve was stable over the
one-year baseline of our observations, changing by less than 0.05 mag.
In the final step, we converted the data from magnitudes to fluxes
\citep{megessier1995}.

The observations from each run were combined into mean light curves
using the linear ephemeris of \citet{mcclintock1986}.  The combined
fluxes and error bars were determined from weighted means of the data,
where the weights were based on the uncertainties assigned each night
from the scatter about the mean for the third field star.  The bin
sizes in the combined light curves are 0.01 in orbital phase for J and
H and 0.05 for K, corresponding to time intervals of 4.65 min and 23.3
min, respectively.  The orbital phases conform to the standard
convention: phase 0 corresponds to inferior conjunction of the K star
%(note that
%this is 180$\arcdeg$ out of phase with respect to the phasing of the
%McClintock \& Remillard ephemeris).  
There is no substantial drift in the phasing of the light curve
relative to the 1986 ephemeris (nor with respect to the refined
ephemeris of Orosz et al.\ 1994; the two orbital solutions differ by
an amount smaller than the bin size of our points): the light curve
minima occur at phases 0 and 0.5, and the maxima at phases 0.25 and
0.75.

\begin{figure}
\plotone{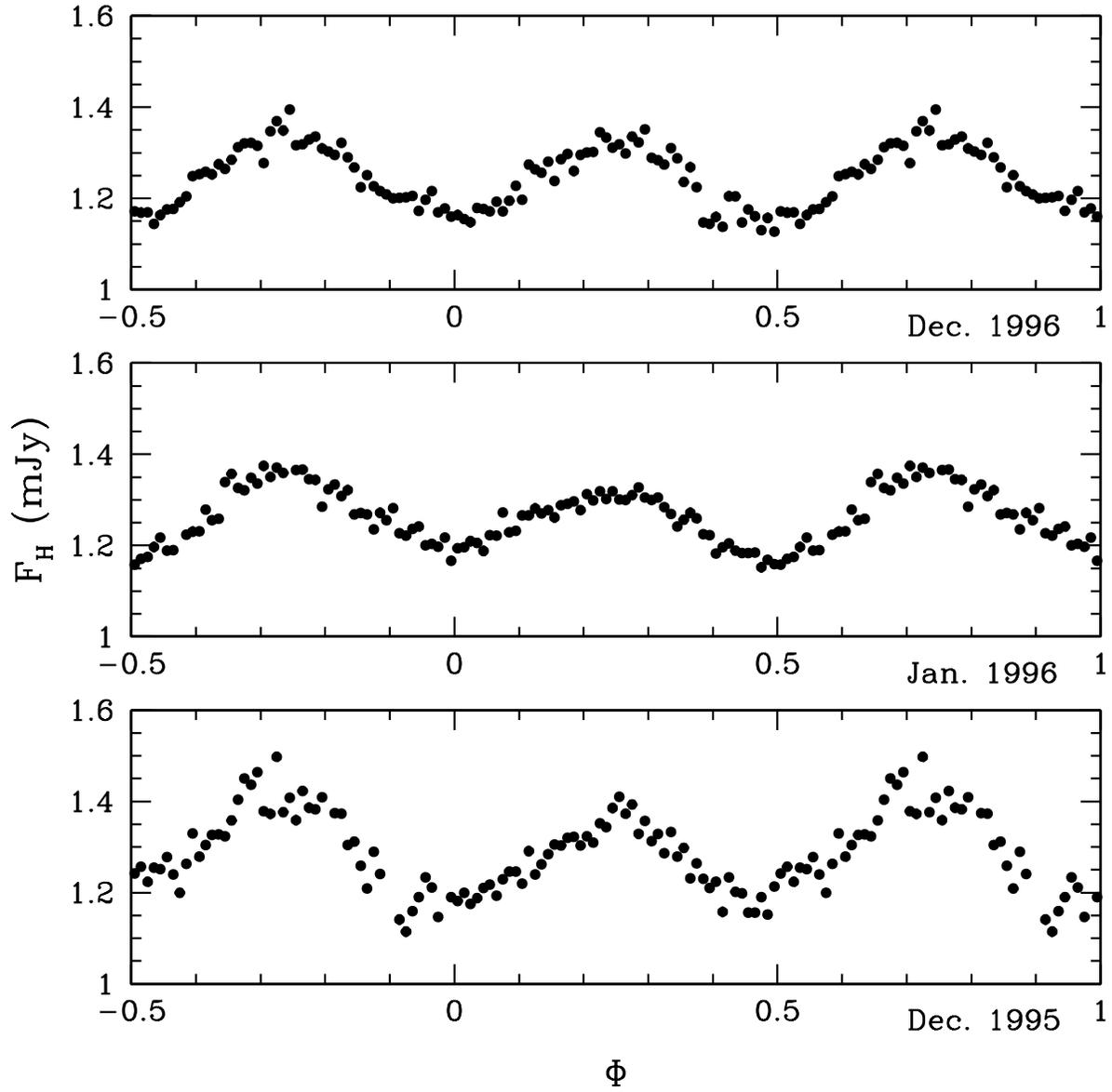}
\figcaption[Froning.fig1.ps]{The 1996 December, 1996 January and 1995
December H-band light curves of A0620--00, plotted over one and a half
cycles in orbital phase.  The data are binned to 0.01 in orbital
phase.  Orbital phase zero corresponds to inferior conjunction of the
mass-losing star.\label{fig_a0620h}}
\end{figure}

\begin{figure}
\plotone{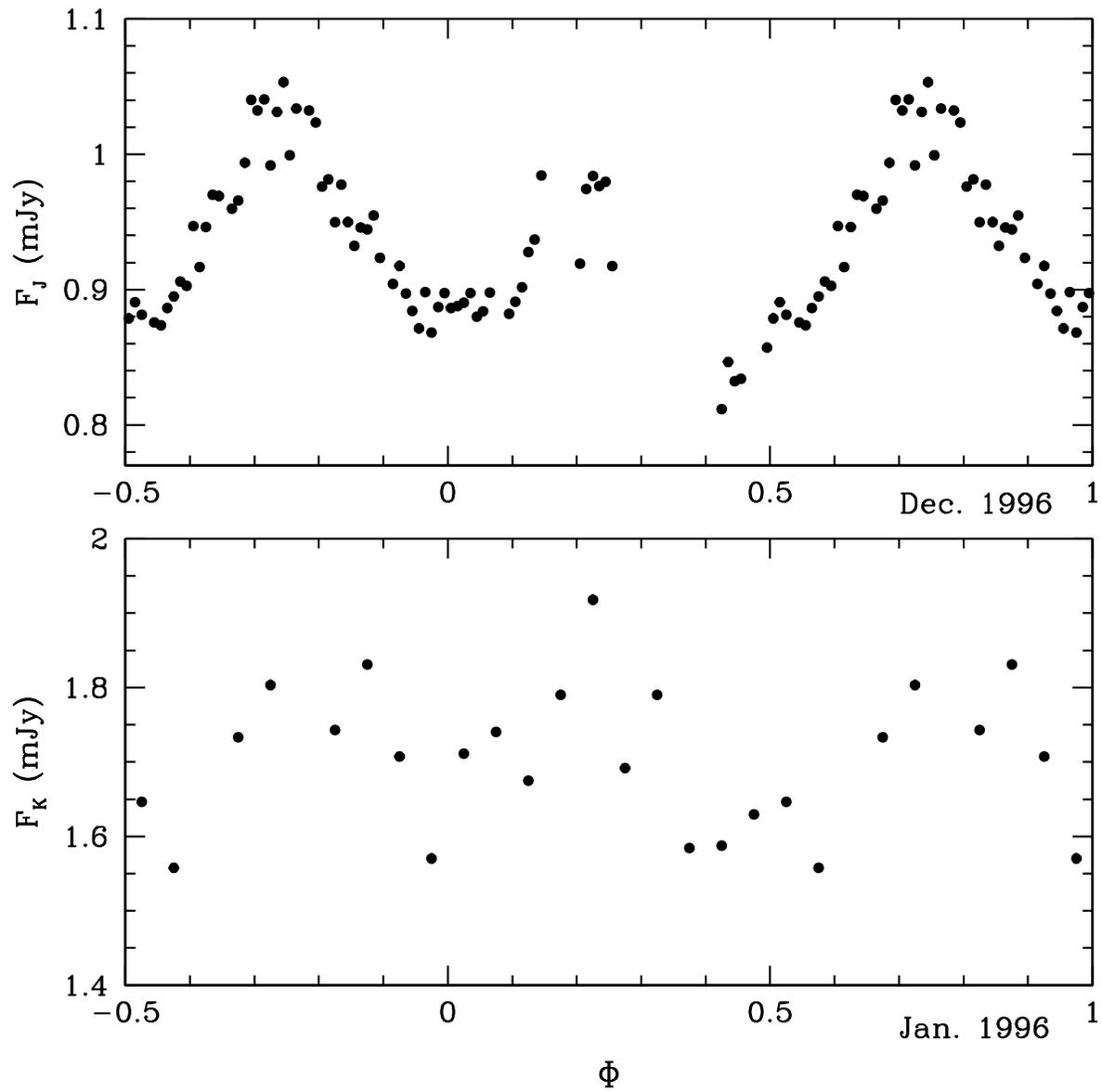}
\figcaption[Froning.fig2.ps]{The 1996 December J-band light curve and the
1996 January K-band light curve of A0620--00, plotted over one and a
half cycles in orbital phase.  The J observations are binned to 0.01
in orbital phase.  The K observations are binned to 0.05 in orbital
phase. Orbital phase zero corresponds to inferior conjunction of the
mass-losing star. \label{fig_a0620jk}}
\end{figure}

The H-band light curves of A0620--00 are shown in
Figure~\ref{fig_a0620h}, and the observations in J from 1996 December
and in K from 1996 January are shown in Figure~\ref{fig_a0620jk}.  The
light curves all show an asymmetric, double-humped modulation.  There
were small fluctuations over the orbit from one observation to the
next, but there was no gross variability in the shape or overall flux
of the H light curve over the one-year course of our observations.
The minimum at $\phi$ = 0.5 is deeper than the primary minimum at
$\phi$ = 0 and the peak at $\phi$ = 0.75 higher than the peak at
$\phi$ = 0.25 in all of the H-band light curves, and the same appears
to be true in the (noisier) K light curve.  The J-band light curve is
too poorly sampled to determine the relative amplitudes of the peaks
and troughs.  The peaks and troughs in the data are smooth, showing no
evidence of the sharp dips seen in B light curves circa 1986--1989
\citep{bartolini1990,haswell1993,haswell1996}.  The shape of the light
curves is very similar to that of a K-band light curve obtained in
1990 by \citet{shahbaz1994}.  The phasing of the light curve peaks and
troughs and the sense of the asymmetry in the peaks is the same in
both observations.  Our mean K and (J--K) colors are also close to the
colors they found.

The shapes of the near-infrared light curves of A0620--00 are
generally consistent with ellipsoidal modulation in a system of
moderate to high inclination, but there are two clear deviations from
a pure ellipsoidal modulation.  First, there are small fluctuations
($\simeq$0.05 -- 0.1~mJy) among the three observations in the
amplitudes of the peaks relative to the troughs, indicating a variable
source of flux in addition to the ellipsoidal modulation.  The second
deviation is the unequal maxima, which an ellipsoidal variation cannot
produce.  The bright spot is a plausible source for this asymmetry.
It is seen in Doppler tomograms of the accretion disk in A0620--00
\citep{marsh1994} and would be expected to boost the peak at $\phi$ =
0.75 relative to the $\phi$ = 0.25 peak, as we observe.  Superhumps
from a precessing, non-circular disk or star spots on the K star could
also cause asymmetries but, as we will show, they are less plausible
sources for the asymmetries in our light curves.

\section{Modeling the Light Curves} 

We modeled the near-infrared light curves of A0620--00 with a
rewritten, updated version of the light curve synthesis code described
in \citet{zhang1986}.  The code simulates the light curves of binary
systems and includes the equipotential geometry, limb and gravity
darkening for the stars, and a limb-darkened, flared accretion disk
with a bright spot on its edge and top surface.  A significant
improvement to the code is the use of specific intensities and
quadratic limb darkening coefficients for cool stars obtained from
fits to model stellar atmospheres of late-type stars (Allard \&
Hauschildt 1995; the fits to the model atmospheres are presented in
Froning 1999b).

We fit three sets of models to the A0620--00 light curves.  In the
first set we assumed that only the K star contributes to the observed
flux.  For the second set we added an accretion disk with a bright
spot on its edge.  The third set was a series of models to estimate
the effect of a non-varying source of diluting flux on the derived
orbital inclinations.  We modeled the H light curves only; the J and K
data are of lower quality and do not warrant detailed modeling.

\subsection{Models Including Only the K Star}  \label{sec_star}

We initially assumed that the K star is the only source of the
observed flux.  We assumed that the asymmetry in the observed light
curve is caused by extra flux added to the peak at $\phi$ = 0.75 and,
to avoid this extra flux, we fit the models to the light curves only
over phases from $\phi$ = 0 to 0.51. We generated model light curves
of the K star, varying each of the parameters that affect the shape of
the ellipsoidal modulation in turn, and fit the light curves to the
observed data by least squares.

The parameters that affect this model are: the orbital inclination,
$i$, the mass ratio, $q$, the temperature of the K star at the pole,
$T_{2}$, the quadratic limb-darkening coefficients and the gravity
darkening coefficient, $\beta$.  We varied the inclination from $i$ =
1$\arcdeg$ -- 89$\arcdeg$ in 1$\arcdeg$ steps and the K star
temperature from $T_{2}$ = 4000 -- 4500~K in increments of 100~K.  The
mass ratio in A0620--00 is known to be $q$ = 0.067 \citep{marsh1994}.
We calculated models for this mass ratio but also for $q$ = 0.056,
0.083 and 0.10 to check the dependence of our results on mass ratio.
The limb darkening was modeled using quadratic limb darkening
coefficients obtained from the model stellar atmospheres discussed
above.  We calculated models for gravity darkening coefficients of
0.05 and 0.08 \citep{sarna1989} and we assumed that the K star fills
its Roche lobe.

\begin{figure}
\plotone{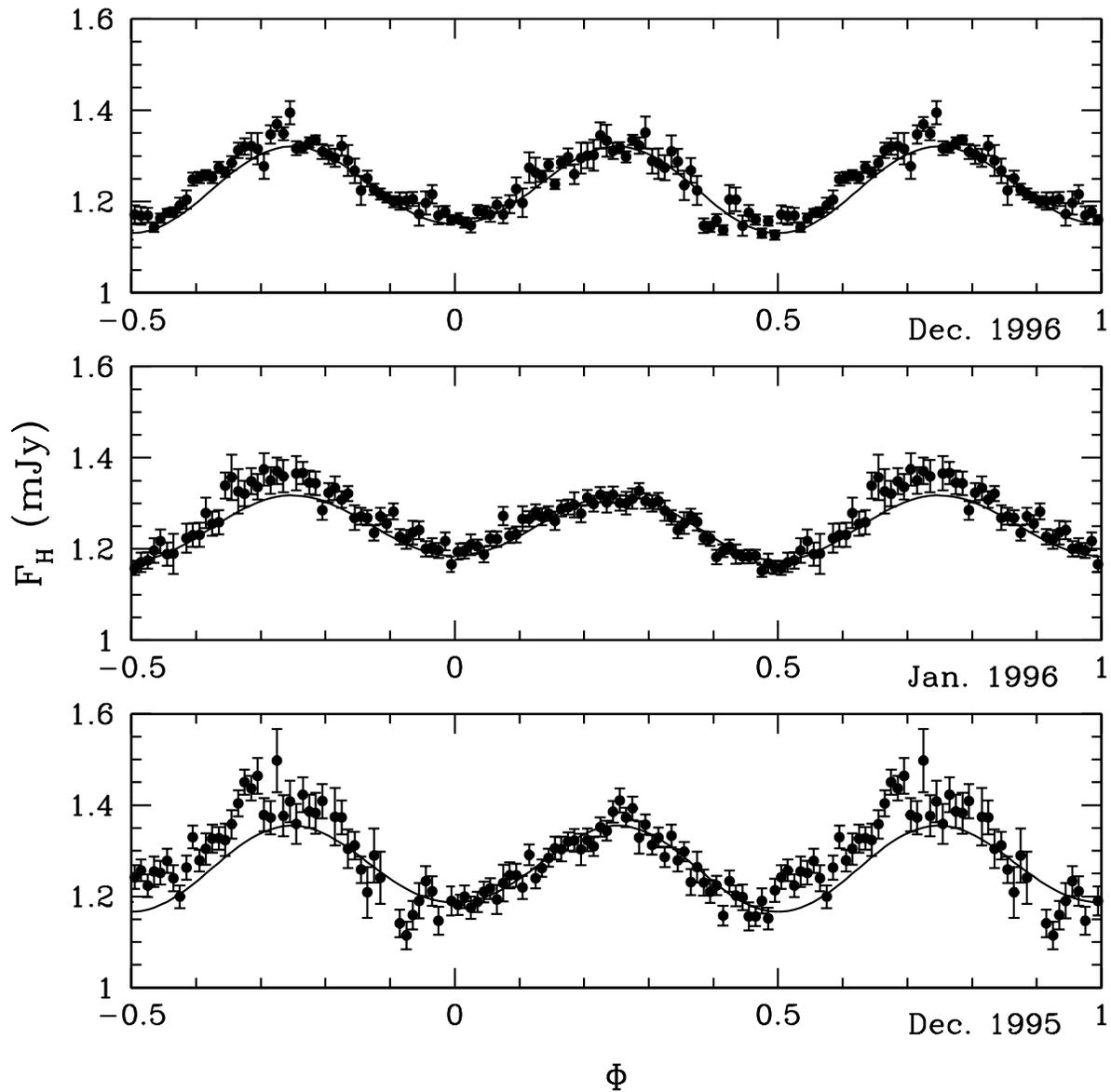}
\figcaption[Froning.fig3.ps]{The H-band light curves of A0620--00 and
the best-fit model including only the K star ellipsoidal variations.
The models were fit to the light curves from $\phi$ = 0 -- 0.51 only,
assuming that the K star is the sole source of flux at those orbital
phases.  The orbital inclinations of the models are $i$ = 44$\arcdeg$,
38$\arcdeg$ and 45$\arcdeg$ for 1996 December, 1996 January and 1995
December, respectively. \label{fig_mod1}}
\end{figure}

Figure~\ref{fig_mod1} shows the best-fit models for each of the H-band
light curves.  The parameters for the models are given in
Table~\ref{tab_fits}. For the 1996 December light curve, the best-fit
model for $q$ = 0.067 has an inclination of $i$ = 44$\arcdeg$; for the
1996 January light curve, $i = 38\arcdeg$ gives the best fit; and for
the 1995 December light curve, models with 43$\arcdeg$ -- 45$\arcdeg$
give good fits.  The models are largely insensitive to variations in
the mass ratio and are only weakly dependent on the temperature of the
K star and the value of the gravity darkening coefficient.  The best
fits were typically obtained for $T_{2}$ = 4100~K and $\beta$ = 0.08.
More important, changes in these parameters had virtually no effect on
the value of the best-fit inclination.  For the 1996 January light
curve, for example, the reduced $\chi^2$ of the best fits for each
combination of the other parameters tested range from $\chi^{2}_{\nu}$
= 1.06 -- 1.58, but the inclinations for these models vary only from
$i$ = 38$\arcdeg$ -- 41$\arcdeg$.  Similarly, the models calculated
for the 1996 December light curve had $\chi^{2}_{\nu}$ between 2.62
and 3.01 and inclinations from $i$ = 45$\arcdeg$ -- 50$\arcdeg$, while
for the 1995 December light curve, $\chi^{2}_{\nu}$ = 1.13 -- 1.18 and
$i$ = 43$\arcdeg$ -- 48$\arcdeg$.

If the K star is the sole contributor to the near-infrared emission
from $\phi$ = 0 -- 0.51, then the orbital inclination in A0620--00 is
$i$ = 38$\arcdeg$ -- 45$\arcdeg$ ($i$ = 38$\arcdeg$ -- 50$\arcdeg$ for
the extreme range of parameter values).  This range is consistent with
the results of \citet{shahbaz1994}, who found a best fit to their K
light curve of $i$ = 37$\arcdeg$ for $q$ = 0.067 and a 90\% confidence
interval of $i \simeq$ 30$\arcdeg$ -- 45$\arcdeg$ for that mass ratio
(their Fig.\ 2).  

The assumption that the modulation of the light curve at these orbital
phases is purely ellipsoidal is not correct, however.  The amplitude
of the $\phi$ = 0.25 peak relative to the light curve minima is
smaller in the 1996 January light curve than in 1995 December or 1996
December light curves.  This leads to a lower value for the orbital
inclination based on the 1996 January data ($i = 38\arcdeg$) and a
range of possible inclinations (38$\arcdeg$ -- 41$\arcdeg$) that does
not overlap those determined from the 1995/1996 December light curves
($i$ = 43$\arcdeg$ -- 50$\arcdeg$).  This change in the amplitude of
the modulation even at $\phi =0.25$ indicates that there is some
source of variable contamination of the ellipsoidal modulation.  It
also gives a measure of the uncertainty in any determination of the
inclination based on just one observation epoch: the inclinations
derived from our three light curves have a range of 7$\arcdeg$.  In
summary, models including only the effect of ellipsoidal variations of
the K star give a lower limit to the inclination of $i \geq 38\arcdeg$
and show that there is a variable distortion of the ellipsoidal
variations even at $\phi =0.25$.

\subsection{Models Including the K Star and an Accretion Disk with a 
Bright Spot} \label{sec_disk}

We next modeled the light curves over the full binary orbit, adding an
accretion disk and a bright spot to the K star.  We have complete
orbital coverage only in the H band, so we cannot constrain system
parameters such as the temperature of the accretion disk and bright
spot with our models.  Rather, our goal was to find simple models with
reasonable parameter values that fit the observed light curves in
order to determine the range of possible values for the orbital
inclination.

Since the derived orbital inclinations do not depend strongly on the
mass ratio, the temperature of the K star, nor the gravity darkening
coefficient, we fixed their values at $q$ = 0.067, $T_{2}$ = 4100~K
and $\beta$ = 0.08.  The inner radius of the accretion disk was fixed
at 0.001~R$_{L_{1}}$ and the outer radius at 0.5~R$_{L_{1}}$, which is
the outer radius in visible light found by \citet{marsh1994}.  To
allow for a beamed bright spot on the disk rim, we set the flare
half-angle of the disk to a small but non-zero value of 1$^{\circ}$.
The bright spot also extends onto the top surface of the disk from
0.45 -- 0.5~R$_{L_{1}}$ \citep{marsh1994}.  Since the bright spot is
not eclipsed and since the disk has only a small flare, the spot
component on the top of the disk merely adds to the constant disk flux
and is otherwise irrelevant.  The bright spot and accretion disk were
each assumed to emit as single-temperature blackbodies, with the
linear limb-darkening coefficients for the accretion disk obtained
from \citet{claret1998}.  Previous observations of A0620--00 in
quiescence have found no evidence for significant irradiation of the K
star by the disk, so we did not include irradiation in our models.

The parameters we varied were: the orbital inclination, $i$, the
temperature of the accretion disk, $T_{disk}$, the temperature of the
bright spot, $T_{spot}$, the azimuthal position of the spot on the
disk, $\phi_{spot}$ and the azimuthal full width of the spot, $\Delta
\phi_{spot}$.  We varied the inclination from $i$ = 1$\arcdeg$ --
89$\arcdeg$, T$_{disk}$ from 2000~K -- 5000~K, T$_{spot}$ from 5000~K
-- 25,000~K, $\phi_{spot}$ from 80$\arcdeg$ -- 115$\arcdeg$ (measured
with respect to the line connecting the centers of the stars and
increasing in the direction of the orbital motion) and $\Delta
\phi_{spot}$ from 5$\arcdeg$ -- 15$\arcdeg$.

\begin{figure}
\plotone{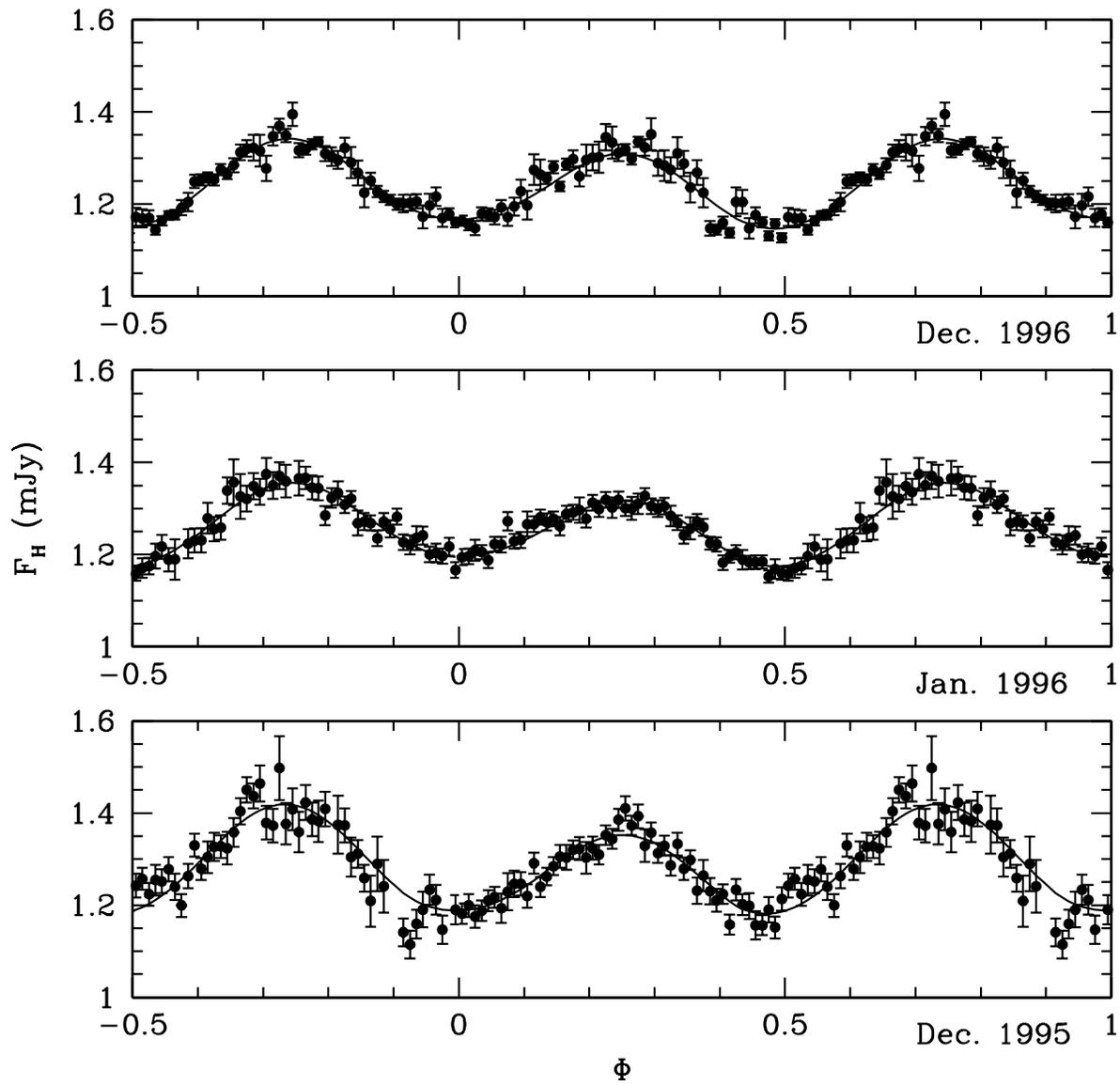}
\figcaption[Froning.fig4.ps]{The H-band light curves of A0620--00 with
an example of a successful K star plus accretion disk and
bright spot model overplotted for each. For 1996 December, the model
shown has $i$ = 74$\arcdeg$ and T$_{disk}$ = 5000~K.  For 1996
January, $i$ = 70$\arcdeg$ and T$_{disk}$ = 5000~K and for 1995
December, $i$ = 53$\arcdeg$ and T$_{disk}$ = 2000~K. \label{fig_mod2}}
\end{figure}

Figure~\ref{fig_mod2} shows the H light curves and the models with the
lowest $\chi^2$ for the parameter values we tested.  The output model
parameters are given in Table~\ref{tab_fits}.  The fits demonstrate
that simple models incorporating ellipsoidal variations from the K
star and a two-component --- constant flux plus a beamed bright spot
--- accretion disk can fully account for the near-infrared light
curves of A0620--00.  We reiterate that the models summarized above
are not intended to provide real constraints on the properties of the
accretion disk (the temperature of the bright spot, for example, is
dependent on its assumed size and on the assumed temperatures of the K
star and the accretion disk).  Moreover, the fits presented above are
definitely not unique.  Virtually any model with $i > 38\arcdeg$ will
fit the light curves equally well if $T_{disk}$ is increased with
inclination.  For example, a model with $i = 44\arcdeg$ and $T_{disk}$
= 2000~K fits the 1996 January light curve as well as the model with
$i = 70\arcdeg$ and $T_{disk}$ = 5000~K shown in
Figure~\ref{fig_mod2}.  

Our light curves show no evidence of an eclipse, which does set an
upper limit to the inclination.  For the assumed disk size, $R_{disk}$
= 0.5~R$_{L_{1}}$, we found that inclinations above 75$\arcdeg$
introduced eclipse features in the models inconsistent with the data.
Marsh, Robinson \& Wood (1994) found a similar limit, $i < 76\arcdeg$,
based on the lack of observed rotational disturbance in the H$\alpha$
emission line.  Our agreement with their upper limit is reassuring but
not unexpected, as we used their values for the mass ratio and
accretion disk radius in our models.

The sense of the asymmetry of the H-band light curve and the orbital
phasing of the light curve peaks was stable over the one year covered
by our observations.  This long-term stability argues against a
precessing, non-circular disk or star spots as sources of the
asymmetry in the light curve peaks, as both are likely to produce a
variable asymmetry.  When the asymmetry is modeled by a bright spot on
the edge of the disk, both the phase and the extent of the bright spot
are the same in all three H-band light curves and, with less
confidence, in the K-band light curve: $\phi_{spot} = 100^\circ -
110^\circ$ and $\Delta \phi_{spot} = 5^\circ - 10^\circ$.  These
results strongly support identifying the bright spot as the source of
the extra flux at $\phi$ = 0.75.  The location of the bright spot in
our models is not the same as the location of the spot in the Doppler
maps of \citet{marsh1994}, who found $\phi_{spot} \simeq 55\arcdeg$
(their Figure 7).  The position of a bright spot as seen in optical
line emission will not necessarily coincide with the location of peak
spot emission in the near-infrared (see, e.g., Littlefair et al.\
2000, Froning et al.\ 1999) ; the difference between the two suggests
that the near-infrared bright spot emission in A0620--00 originates
downstream from the initial mass stream impact point.

\subsection{Models with a K Star and a Constant Extra Flux.} \label{sec_third}

Since there is a direct tradeoff between extra constant flux from the
accretion disk (or any other source) and the orbital inclination
inferred from the amplitude of the ellipsoidal variations, we ran a
final set of models to determine how the inclination changes as disk
flux is added to the light curve.  To do this, we fit a K star model
(T$_{2}$ = 4100~K, $\beta$ = 0.08) to the 1996 January data and added
increasing amounts of disk flux, specified in the light curve
synthesis program as a fraction of the total light curve flux at
$\phi$ = 0.25.  Table~\ref{tab_third} shows the derived orbital
inclination as the disk contribution increases.  For no disk
contribution, we recover the 38$\arcdeg$ inclination found in
Section~\ref{sec_star}.  To allow orbital inclinations approaching
75$\arcdeg$ -- the upper limit set by the lack of eclipses -- the disk
would need to be contribute 55\% of the H-band flux.
Table~\ref{tab_third} also shows that if the accretion disk
contribution in the H-band is low, large fluctuations in the disk flux
are needed to explain the change in the amplitude of the $\phi$ = 0.25
peak discussed in Section~\ref{sec_star}: the 6--7$\arcdeg$ difference
in the inclinations of our secondary star model fits to the three H
light curves implies fluctuations of 20--25\% in the contaminating
flux at the $\phi$ = 0.25 peak.

\section{The Mass of the Compact Star}

Based on our fits to the H-band light curves, we can constrain the
inclination in A0620--00 to $38\arcdeg \leq i \leq 75\arcdeg$.  The
upper limit on the inclination is quite strict, as light curve models
with $i \geq 75\arcdeg$ show both primary and secondary eclipse
features not seen in the data. The lower limit on the inclination is
also a fairly strict limit.  The amplitude of the light curve
modulation was larger in 1996 December and 1995 December than in 1996
January, indicating that the 38$\arcdeg$ lower limit on the
inclination derived from the latter is a probable underestimate of the
true binary inclination caused by dilution of the ellipsoidal
modulation by an accretion disk flux component.

From determinations of the orbital period, mass ratio and radial
velocity semi-amplitude of the K star, \cite{marsh1994} derived a mass
$M_{1} = (3.09\pm0.09)\ \sin^{-3}i$~\msun\ for the compact star in
A0620-00.  Combined with our limits on the inclination, this limits
the mass of the compact star to lie in the range $3.3 \leq M_{1} \leq
13.6$~\msun.  The lower limit is close to, but remains larger than,
the maximum mass of a uniformly rotating neutron star with the
stiffest equation of state, $\sim$3.2~\msun\ \citep{friedman1986}, and
it is well above the maximum mass of a nonrotating neutron star, 1.8
-- 2.5~\msun\ \citep{akmal1998}.

\citet{haswell1993} identified a sharp dip in their UBVR light curves
of A0620-00 as an eclipse of the K star by a large accretion disk (at
or near the maximum radius for a circular disk inside the Roche lobe
around the compact star).  If this identification is correct, the
inclination of A0620-00 is $65\fdg75 \leq i \leq 73\fdg5$ and the mass
of the compact star is $3.40 \leq M_{1} \leq 4.20$~\msun (using the
inclination range found by Haswell et al.\ 1993 for $q$ = 0.067 and
masses from the Marsh, Robinson \& Wood 1994 equation for $M_{1}$
given above).  \citet{johnston1989} also invoked a large accretion
disk to model 1986 spectroscopic observations of A0620--00.
\citet{haswell1996} explained the large disk and the grazing eclipse
in the context of a non-circular, precessing accretion disk model for
A0620--00: as the orientation of the non-circular disk changes over
the long ($>>P_{orb}$) precession period, eclipses appear and
disappear and superhumps move through the light curve, causing changes
in the shape of the light curves as was, indeed, seen from 1981 --
1989 in A0620--00.

\citet{leibowitz1998} found that A0620--00 exhibited slow fluctuations
in its mean optical brightness between 1991 and 1995 with a
peak-to-peak amplitude of 0.3 -- 0.4 mag in the R band.  They did not
find any periodicities in the fluctuations.  Specifically, they found
no evidence of a superhump period nor of a beat period between the
superhump and orbital periods.  The shape of the orbital light curve
varied somewhat with brightness, but the sense of the asymmetry
between the two peaks never reversed.  There is, therefore, no
evidence for current or recent superhumps in A0620--00.
\citet{haswell1996} concluded that, after 1989 the accretion disk had
shrunk below the radius necessary to trigger the tidal interactions
that drive superhumps.

The most serious challenge to the superhump model comes from
\citet{marsh1994}, who re-analyzed the 1986 observations of
\citet{johnston1989} and concluded that both the 1986 and the 1991/92
data sets---which straddle the observations of
\citet{haswell1993}---are consistent with a smaller accretion disk,
$R_{disk}$ = 0.5 -- 0.6 $R_{L_{1}}$, one too small to drive
superhumps.  They note, however, that their accretion disk radius is
based on the radius of H$\alpha$ emission and could be smaller than
the radius of the optically thick accretion disk used by
\citet{haswell1993}.

\citet{shahbaz1994} also narrowly constrained the derived inclination
for A0620--00 to $30\arcdeg \leq i \leq 45\arcdeg$,
%(90\% confidence limits for $q$ = 0.067 from their Fig.\ 2),
corresponding to $8.49 \leq M_{1} \leq 25.4$~\msun.  Their best-fit
value for $q$ = 0.067 was $i$ = 37$\arcdeg$, which corresponds to
$M_{1}$ = 14.2~\msun.  Their analysis was predicated on the assumption
that the accretion disk does not contaminate the K-band flux from the
K star.  To test this, \citet{shahbaz1999} obtained a K-band spectrum
of A0620--00 to which they fit scaled template spectra of stars of
known spectral type.  From this, they concluded that the K star
provides 75$\pm$17\% of the K-band flux.  They noted that a 27\%
accretion disk contribution (their maximum likely disk fraction) would
increase the minimum inclination in A0620--00 by 7$\arcdeg$ and
decrease the mass of the compact star by 3.6~\msun.  If the accretion
disk does contribute $\sim$25\% of the near-infrared flux, our three
observations would indicate an inclination of $i = 46\arcdeg$ --
53$\arcdeg$ and a corresponding mass of $5.9 < M_{1} < 8.5$~\msun.

To determine the contribution of the disk to the infrared spectrum,
\citet{shahbaz1999} fit the template spectra to just the $\lambda2.29\
\mu$m $^{12}$CO(2,0) bandhead.  The use of this feature to estimate
the disk contribution is problematic for several reasons.  First, the
CO molecular line strengths are both temperature and gravity dependent
(e.g., Kleinman \& Hall 1986).  While the spectrum of the K star in
A0620--00 is clearly inconsistent with that of a giant star
(luminosity class III), its effective gravity could still be
significantly lower than the gravity of a main-sequence star
\citep{oke1977,murdin1980}.  \citet{shahbaz1999} used only dwarf stars
for their template spectra, so their results could not distinguish
gravity dependent changes in the $\lambda2.29\ \mu$m $^{12}$CO(2,0)
line strength.  Second, observations of cataclysmic variables have
shown that in some systems, $^{12}$CO absorption is weaker relative to
the strengths of the atomic absorption lines than expected for their
donor star spectral types \citep{harrison1999}.  In addition, the
ratio of the $^{12}$CO bandhead equivalent width to those of the
atomic lines can vary with time \citep{ramseyer1993}.  The reason for
these abnormalities in the CO absorption line strengths is not yet
understood.  Thus, using the strength of the $\lambda2.29\ \mu$m
$^{12}$CO(2,0) bandhead to measure the contribution of the disk to the
infrared flux gives unreliable results.  The measurements by
\citet{shahbaz1999} do not settle the question of how much flux the
disk contributes to the light curve of A0620-00 at infrared
wavelengths.  The orbital inclination could, therefore, be
significantly higher than the lower limit set by the ellipsoidal
variations ($i > 38\arcdeg$) and the mass of the compact star much
less than the upper limit ($M_{1} < 13.6$~\msun).

\section{Conclusions}

1. The infrared light curves of the black hole binary A0620--00 show
an asymmetric, double-humped modulation with extra emission in the
peak at phase $\phi$ = 0.75.  There were no gross changes in the
morphology of the light curves over a one year period from 1995
December to 1996 December.  The mean infrared colors and the shape of
the light curve are also the same as observed in 1990 January.  There
were no sharp dips in the light curve nor reversals of the asymmetry
between the two humps as were seen in observations circa 1986 -- 1989.

2. The light curves are consistent with ellipsoidal variations from
the K star plus beamed flux from a spot on the accretion disk.  A
precessing disk is an unlikely source of light curve modulation during
the observation period.  Star spots are also ruled out unless the spot
is fixed in location and size on the K star surface.
%The similarity of the light curves in morphology to others obtained
%post-1989 suggest that ellipsoidal variations are and have been the
%dominant light curve modulation since that time.

3. Based on fits to the lower peak in the light curve (between $\phi$
= 0 and 0.51), the minimum inclination in A0620--00 is $i \geq
38\arcdeg$.  From the absence of eclipse features in the light curves,
models including an accretion disk of radius $R_{d} = 0.5\ R_{L_{1}}$
give an upper limit to the inclination of $i \leq 75\arcdeg$.  The
mass of the compact star in A0620--00 is $3.3 < M_{1} < 13.6$~\msun.

4.  Ellipsoidal variations provide only a lower limit to the
inclination of non-eclipsing binary systems when the contribution of
the accretion disk is unknown.  A previous attempt to determine the
relative contribution of the accretion disk and K star to the
near-infrared flux used the strength of the $\lambda2.29\ \mu$m
$^{12}$CO(2,0) bandhead.  This method is not reliable. 

\acknowledgments{We thank Erik Fierce for observing A0620--00 in 1996
December.}

\pagebreak

\clearpage
\begin{deluxetable}{lcc}
\tablecaption{Observations of A0620--00 \label{tab_logobs}}
\tablewidth{0pt}
\tablecolumns{3}
\tablehead{
\colhead{Date (UT)} & \colhead{Filter(s)} & 
\colhead{Total Exposure Time (ksec)} }
\startdata
1995 Dec 15     &       H               & 3.72              \\
1995 Dec 16     &       H               & 3.46              \\
1995 Dec 17     &       H               & 1.92              \\
1996 Jan 27     &       H,K             & 3.60, 0.25        \\
1996 Jan 28     &       H               & 0.70              \\
1996 Jan 29     &       H,K             & 4.40, 0.35        \\
1996 Dec 7      &       H               & 2.96              \\
1996 Dec 8      &       J,H             & 1.16, 0.72        \\
1996 Dec 9      &       J,H             & 1.34, 1.48        \\
1996 Dec 10     &       J,H             & 1.16, 1.40        \\
1996 Dec 12     &       J,H             & 0.20, 3.56        \\ 
\enddata
\end{deluxetable}

\clearpage
\begin{deluxetable}{ccccc}
\tablecaption{Reference Stars in A0620--00 Field \label{tab_stars}}
\tablewidth{0pt}
\tablecolumns{5}
\tablehead{
\colhead{Star} & \colhead{Position\tablenotemark{a}} & \colhead{J} &
\colhead{H} & \colhead{K}}
\startdata
Star 1 & 18$\arcsec$N, 18$\arcsec$W & 13.4$\pm$0.1 & 12.92$\pm$0.02 & 12.78$\pm$0.03 \\
Star 2 & 19$\arcsec$N, 5$\arcsec$W  & 15.0$\pm$0.1 & 14.34$\pm$0.02 & 14.06$\pm$0.03 \\
Star 3 & 37$\arcsec$N, 14$\arcsec$E & 15.4$\pm$0.1 & 14.79$\pm$0.02 & 14.57$\pm$0.03 \\
\enddata
\tablenotetext{a}{Relative to A0620--00.}
\tablecomments{The error bars given are the uncertainties in the flux 
calibration only.  In J and H, this is the dominant 
uncertainty, but in K, where little data was acquired, the standard 
deviation about the mean for each star's measurements is of order the 
flux calibration uncertainty.}
\end{deluxetable}

\clearpage
\begin{deluxetable}{lcc}
\tablecaption{NIR Colors of A0620--00 \label{tab_colors}}
\tablewidth{0pt}
\tablecolumns{3}
\tablehead{
\colhead{Date} & \colhead{Filter} & \colhead{Mean color} }
\startdata
1995 December   &       H       &       14.80$\pm$0.02  \\
1996 January    &       H       &       14.83$\pm$0.02  \\
1996 January    &       K       &       14.49$\pm$0.03  \\
1996 December   &       J       &       15.6$\pm$0.1    \\
1996 December   &       H       &       14.84$\pm$0.02  \\ 
\enddata
\tablecomments{The error bars shown are the uncertainties in the flux
calibration only.}
\end{deluxetable}

\clearpage
\begin{deluxetable}{lcccccc}
\tablecaption{Parameters of best fit light curve models. \label{tab_fits}}
\tablewidth{0pt}
%\rotate
\tablecolumns{7}
\tablehead{
\colhead{Light Curve} & \colhead{$\chi^{2}_{\nu}$} & \colhead{$i$} &
\colhead{T$_{disk}$} & \colhead{T$_{spot}$} & \colhead{$\phi_{spot}$} &
\colhead{$\Delta\phi_{spot}$} \\
\cline{1-7} \\
\multicolumn{7}{c}{Secondary star only} }
\startdata
1996 December & 2.62 & 44$\arcdeg$ & \nodata & \nodata & \nodata & \nodata \\
1996 January & 1.06 & 38$\arcdeg$ & \nodata & \nodata & \nodata & \nodata \\
1995 December & 1.13 & 43$\arcdeg$ -- 45$\arcdeg$ & \nodata & \nodata & \nodata & \nodata \\
\cutinhead{Secondary star plus accretion disk and bright spot}
1996 December & 1.88 & 74$\arcdeg$ & 5000~K & 25,000~K & 100$\arcdeg$ & 5$\arcdeg$ \\
1996 January & 0.83 & 70$\arcdeg$ & 5000~K & 20,000~K & 100$\arcdeg$ & 10$\arcdeg$ \\
1995 December & 1.32 & 53$\arcdeg$ & 2000~K & 25,000~K & 115$\arcdeg$ & 10$\arcdeg$ \\
\enddata
\tablecomments{For all models listed, $q$ = 0.067, T$_{2}$ = 4100~K, 
$\beta$ = 0.08.
For the models including an accretion disk, the disk extends from 
0.001~R$_{L_{1}}$ to 0.5~R$_{L_{1}}$.  The radial extent of the bright spot 
is 0.45 -- 0.5|R$_{L_{1}}$, and the disk has a flare half-angle of 1$\arcdeg$.}
\end{deluxetable}

\clearpage
\begin{deluxetable}{lllll}
\tablecaption{Best fit inclination vs. fractional disk contribution 
for 1996 January lightcurve. \label{tab_third}}
\tablewidth{0pt}
\tablecolumns{5}
\tablehead{
\colhead{$i$ ($\arcdeg$)} & \colhead{$f_{disk}$} & \colhead{} &
\colhead{$i$ ($\arcdeg$)} & \colhead{$f_{disk}$} }
\startdata
38 & 0 & & 54 & 0.4 	\\
39 & 0.05 & & 58 & 0.45	\\
41 & 0.1 & & 64 & 0.5	\\
42 & 0.15 & & 72 & 0.55	\\
44 & 0.2 & & 89 & 0.6	\\
46 & 0.25 & & 89 & 0.65	\\
48 & 0.3 & & 89 & 0.7	\\
51 & 0.35		\\
\enddata
\end{deluxetable}

\newpage
\pagestyle{empty}

\newpage
\pagestyle{empty}

\newpage
\pagestyle{empty}

\newpage
\pagestyle{empty}

\end{document}